\documentclass[useAMS,usenatbib]{mn2e}
\usepackage{natbib}
\usepackage{times}
\usepackage{graphicx}
\usepackage{pdfpages}
\usepackage{pdflscape}

\newcommand{\lae}{\lower 2pt \hbox{$\, \buildrel {\scriptstyle <}\over {\scriptstyle
\sim}\,$}}
\newcommand{\gae}{\lower 2pt \hbox{$\, \buildrel {\scriptstyle >}\over {\scriptstyle\sim}\,$}}

\begin{document}

\title{Discovery of a Long-Lived, High Amplitude Dusty Infrared Transient} 
\date{16 May, 2016}
\pubyear{0000} \volume{000} \pagerange{1}

\author[C.~T. Britt et al.]{C.~T. Britt,$^{1\ast}$ T.~J. Maccarone,$^{1}$ J.~D. Green,$^{2,3}$ P.~G. Jonker,$^{4,12}$ M.~A.~P. Torres,$^{4,5}$ R.~I. Hynes,$^{6}$ 
\newauthor
J. Strader,$^{7}$ L. Chomiuk,$^{7}$ R. Salinas,$^{7}$ P. Lucas,$^{8}$ C. Contreras Pe{\~n}a,$^{8}$ R. Kurtev,$^{9}$ C. Heinke,$^{10}$ 
\newauthor
L. Smith,$^{8}$ N.~J. Wright,$^{8}$ C. Johnson,$^{6}$ D. Steeghs,$^{11}$ G. Nelemans$^{12}$\\
\\
\normalsize{$^{1}$Texas Tech University, Department of Physics, Box 41051 Lubbock, TX 79409-1051, USA}\\
\normalsize{$^{2}$University of Texas at Austin, Department of Astronomy, 2515 Speedway, Stop C1400, Austin, TX 78712-1205}\\
\normalsize{$^{3}$Space Telescope Science Institute, 3700 San Martin Drive, Baltimore, MD, USA}\\
\normalsize{$^{4}$SRON, Netherlands Institute for Space Research, Sorbonnelaan 2, 3584 CA, Utrecht, The Netherlands}\\
\normalsize{$^{5}$European Southern Observatory, Alonso de C{\'o}rdova 3107, Vitacura, Casilla 19001, Santiago de Chile, Chile}\\
\normalsize{$^{6}$Louisiana State University, Department of Physics and Astronomy, Baton Rouge, LA 70803-4001, USA}\\
\normalsize{$^{7}$ Dept. of Physics and Astronomy, Michigan State University, East Lansing, MI 48824, USA}\\
\normalsize{$^{8}$Centre for Astrophysics Research, University of Hertfordshire, College Lane, Hatfield AL10 9AB, United Kingdom}\\
\normalsize{$^{9}$Instituto de F{\'i}sica y Astronom{\'i}a, Universidad de Valpara{\'i}so, Ave. Gran Breta{\~n}a 1111, Valpara{\'i}so, Chile}\\
\normalsize{$^{10}$University of Alberta, Physics Dept., CCIS 4-183, Edmonton, AB, T6G 2E1, Canada}\\
\normalsize{$^{11}$Astronomy and Astrophysics, Dept. of Physics, University of Warwick, Coventry, CV4 7AL, United Kingdom}\\
\normalsize{$^{12}$Dept. of Astrophysics/IMAPP, Radboud University Nijmegen, P.O. Box 9010, 6500 GL, Nijmegen, The Netherlands}\\
\normalsize{$^\ast$To whom correspondence should be addressed; E-mail: christopher.britt@ttu.edu}
}

\maketitle

\begin{abstract}

We report the detection of an infrared selected transient which has lasted at least 5 years, first identified by a large mid-infrared and optical outburst from a faint X-ray source detected with the Chandra X-ray Observatory. In this paper we rule out several scenarios for the cause of this outburst, including a classical nova, a luminous red nova, AGN flaring, a stellar merger, and intermediate luminosity optical transients, and interpret this transient as the result of a Young Stellar Object (YSO) of at least solar mass accreting material from the remains of the dusty envelope from which it formed, in isolation from either a dense complex of cold gas or massive star formation. This object does not fit neatly into other existing categories of large outbursts of YSOs (FU Orionis types) which may be a result of the object's mass, age, and environment. It is also possible that this object is a new type of transient unrelated to YSOs.

\end{abstract}

\section{Introduction}

Transient outbursts offer a unique window into astrophysics by giving astronomers access to physical changes on human timescales. Every class of transient has been of utmost importance in understanding all areas of astronomy, from cosmology to accretion. As wide-field, time-domain surveys grow, the number of very rare events that are detected is growing correspondingly. As new transients are discovered, physical interpretation will often begin with comparisons to existing classes of transient. 

Large amplitude ($\ge 6$ magnitudes) transients come from a number of physical scenarios. Classical novae (CNe) are episodes of runaway fusion of hydrogen on the surfaces of white dwarfs accreting material from a companion \citep{Gehrz98}. These are extreme events in the optical and infrared, reaching $-7<M_V<-9$ \citep{Yaron05}, lasting for tens or hundreds of days and sometimes producing large quantities of dust \citep{Strope10}. In these thermonuclear explosions, mass is ejected from the binary, with the amount dependent upon the mass of the white dwarf and the rate of accretion; CNe requiring less total hydrogen to set off produce less ejecta \citep{Yaron05}. The source of material can be any companion star, from wind feeding by giants (Symbiotic Novae) to Roche-Lobe overflow from Main Sequence donors in Cataclysmic Variables (CVs) to hydrogen deficient donors leading to helium novae \citep{Ashok03}. 

Another class of large amplitude Galactic transient is accretion episodes in Young Stellar Objects (YSOs). During low mass star formation, infalling material forms an accretion disk which serves to transfer angular momentum allowing matter to fall onto the surface of the YSO. Instabilities in the disk can trigger an episode of rapid mass accretion, in which the forming star will gain up to $\approx\,0.02$\,M$_{\odot}$ in an outburst lasting many years \citep{Herbig77,Hartmann96,Miller11}. Infalling envelopes are typically on the scale of 1000-10000 AU and associated with young, less evolved objects $<10^{6}$ years old \citep{Hartmann96,Sandell01,Evans09}. The mid-infrared continuum is associated with reprocessed light from the circumstellar disk and envelope when present \citep{Kenyon91}. The outburst could be caused by a thermal disk instability \citep{Bell95}, or by an interaction with another star which perturbs the disk in a gravitational instability \citep{Bonnell92}. 

While CNe and accretion disk instability events recur, some rarer single-event large amplitude transients can also be discovered by wide-field surveys such as the Vista Variables in Via Lactea survey \citep[VVV,][]{Minniti10}, the Optical Gravitational Lensing Experiment \citep[OGLE,][]{Udalski08}, the Wide-field Infrared Survey Explorer \citep[WISE,][]{Wright10}, and the Galactic Legacy Infrared Midplane Survey Extraordinaire \citep[GLIMPSE,][]{Benjamin03,Churchwell09}. Dramatic, rare variables that such surveys have either discovered or revealed the nature of include stellar mergers, where a binary enters into an unstable regime of mass transfer forming a common envelope, part of which is ejected as the stellar cores inspiral and merge \citep{Tylenda11}. The ejection of a planetary nebula (PN) could also be caught but such surveys.

Some transient classes are based on phenomenology of only a handful of objects, such as Young Stellar Objects (YSOs) undergoing large accretion outbursts \citep{Hartmann96}. These are therefore unlikely to span the full variety of phenomena possible of a particular class of physical system \citep{Contreras16a,Contreras16b}. This is true for any small sample size, but is particularly the case when the method of selection changes, as is the case here. Instead of taking a purely phenomenological approach to classifying the transient reported herein, we will consider the physical mechanisms underlying each class in order to determine if it is possible to produce the observed characteristics.

In the Galactic Bulge Survey \citep[GBS,][]{Jonker11,Jonker14}, an X-ray
survey of part of the Galactic Bulge conducted with the Chandra X-ray
Observatory, we discovered a peculiar transient, CXOGBS J173643.8-282122 (hereafter CX330), where we will show in Sections \ref{sec:interp} and \ref{sec:disc} that the transient class which best fits the data presented in Section \ref{sec:data} is an accretion event in a YSO.

\section{Multiwavelength Data Sources and Methods}

While the source was discovered in optical and infrared wavelengths while already in outburst, archival data in multiple wavelengths covers the region in the years before the outburst begins. In addition, the GBS collaboration took initial photometry of the Bulge region prior to the X-ray observations in order to identify H$\alpha$ excess sources and provide a baseline for spectroscopic follow-up \citep{Wevers16}. In addition to these early observations, we have lightcurves and magnitudes in multiple colors from several other surveys from after the outburst began, and have both optical and near-infared (NIR) spectroscopy of CX330 during outburst.

\subsection{Pre-Outburst Photometry}

We use archival data from surveys of the region which are publicly available to constrain the brightness of CX330 prior to outburst, as well as a set of initial optical observations for the GBS. Several surveys cover the region to depths greater than the post-outburst brightness of CX330, yielding useful constraints on the outburst proginitor and time frame.

\subsubsection{2MASS}

The 2 Micron All Sky Survey \citep[2MASS,][]{Skrutskie06} was a ground based survey using two 1.3 meter telescopes at Mount Hopkins, AZ and Cerro Tololo, Chile. Observations were taken in $JHK$ filters. The survey depth varies with the crowding of the field. While 2MASS reaches quoted sensitivities of $K=14.3$, in the crowded neighborhood of CX330, we find that the cutoff sensitivity is closer to $K=13.0$. The field was observed on July 2nd, 1998. No counterpart to CX330 appears in 2MASS catalogs or is visible in the images themselves.

\subsubsection{Spitzer Space Telescope: GLIMPSE \& MIPSGAL}

{\it Spitzer} conducted surveys of the Galactic Plane and Galactic Bulge in near and mid-infrared wavelengths. GLIMPSE was taken in the 4 IRAC filters, with effective wavelengths of $3.6\mu$m, $4.5\mu$m, $5.8\mu$m, and $8.0\mu$m. Each region of the Plane and Bulge is covered under a different GLIMPSE survey. CX330 lies in the region covered by GLIMPSE-II, which was visited 3 times with exposures of 1.2\,s \citep{Churchwell09}. MIPSGAL covered the same region with effective wavelengths of $24\mu$m and $70\mu$m with the purpose, in part, of identifying all high mass ($M>5M_{\odot}$) protostars in the inner Milky Way disk \citep{Rieke04,Carey09}. The point source catalogs for GLIMPSE and MIPSGAL are publicly available\footnote{http://irsa.ipac.caltech.edu/cgi-bin/Gator/nph-scan?mission=irsa\&submit=Select\&projshort=SPITZER}. We find that in the neighborhood of CX330, the GLIMPSE point source catalog is complete to magnitude $[3.6]<13.5, [4.5]<13.5, [5.8]<12.5, [8.0]<12.$, and that MIPSGAL reach a depth of $[24]<8.6$. 

\subsubsection{GBS Initial Optical Observations}

Initial optical observations for the GBS were taken in 2006 in Sloan $r',i'$, and H$\alpha$ filters on the Blanco 4m Telescope at CTIO with the Mosaic-II instrument, using an exposure time of 2 minutes. The catalog of sources in the 12 square degree region is complete to $r'=20.2$ and $i'=19.2$, while the mean $5\sigma$ depth is $r'=22.5$ and $i'=21.1$ \citep{Wevers16}. Data reduction was carried out using a pipeline created by the Cambridge Astronomical Survey Unit (CASU) \citep{GonzalezSolares08}. 

\subsection{Post-Outburst Photometry, Astrometry, and Spectroscopy}

\subsubsection{GBS Optical Variability Survey}

We acquired 8 nights of photometry, from July 8th-
15th 2010, with the Blanco 4.0 meter telescope at the
Cerro Tololo Inter-American Observatory (CTIO). Using
the Mosaic-II instrument, we observed the 9 square degree area
containing the X-ray sources identified by the first GBS X-ray observations
 \citep{Jonker11} and which contain CX330. The remaining southern GBS sources \citep{Jonker14} were covered in 2013 with DECam observations. Observations were made in Sloan $r'$ with an exposure time of 120\,s. Data reduction, matching variable counterparts to X-ray sources, and other optical variability results are described in detail in \citet{Britt14}. 

\subsubsection{VVV}

The VVV Survey is a time-domain survey of the Galactic Plane and Bulge in 2MASS filters $ZYJHK$ with the VIRCAM instrument on the 4.1m VISTA telescope located at the Paranal Observatory in Chile \citep{Minniti10}, beginning in February, 2010 and having concluded the full initial survey in October, 2015. CX330 is on VVV tile b361 at the start of VVV observations. To estimate line of sight reddening to the Galactic Bulge, we use VVV extinction maps produced by \citet{Gonzalez12}. The $K_{s}$ magnitudes of CX330 were obtained from the tile catalogues produced by the standard CASU pipeline for the VISTA/VIRCAM data. Due to the brightness of the object, the first five epochs of $K_{s}$ data were saturated. To obtain reliable photometry, we measured the flux of the star in a ring with an inner radius of $0.7^{"}$ and outer radius of $1.4^{"}$, thus avoiding the saturated inner core. An aperture correction was derived from non-saturated  stars ($12<$$K_{s}<13.5$ magnitudes) found within $1^{'}$ of CX330. This procedure yields uncertainties of $\sim0.3$ magnitudes. VVV counterparts of GBS X-ray sources are treated in detail in \citet{Greiss14}.

\subsubsection{WISE \& NeoWISE}

WISE is a NASA satellite, launched in December, 2009, sensitive in near and mid infrared wavelengths in 4 passbands with effective wavelengths of $3.4\mu$m, $4.6\mu$m, $12\mu$m, $22\mu$m \citep{Wright10}. Before running out of coolant and being deactivated in February, 2011, it completed an all-sky survey including 2 epochs of the region around CX330. In September, 2013, the satellite was reactivated and the NeoWISE survey continued in the two warmer passbands. Point source catalogs, proper motions, and images for both WISE and NeoWISE are publicly available through IRSA\footnote{http://irsa.ipac.caltech.edu/cgi-bin/Gator/nph-scan?projshort=WISE\&mission=irsa}.

\subsubsection{OGLE-IV}

OGLE-IV \citep{Udalski15} is a time-domain optical survey of the Galactic Bulge, Plane and Magellenic Clouds using the 1.3m Warsaw telescope at Las Campanas Observatory, Chile. Variability information in the field including CX330 was taken in the $I$ filter with exposure times of 100 seconds \citep{Udalski12}. CX330 apears in the OGLE-IV field BLG653.19 as object 81200. 

\subsubsection{Goodman Optical Spectroscopy}

An optical spectrum was taken with the Goodman spectrograph on the SOAR 4.1\,m telescope on March 2, 2014 using the 400 l/mm grating centered at 7000\,\AA\ with a $1.03''$ slit in three 10 minute exposures. It was reduced and extracted using standard packages in IRAF\footnote{IRAF is distributed by the National Optical Astronomy Observatory, which is operated by the Association of Universities for Research in Astronomy (AURA) under cooperative agreement with the National Science Foundation.}. Two additional optical spectra were obtained with the SOAR 4.1m telescope using the Goodman Spectrograph on April 7, 2015, one moderate resolution spectrum with the same instrumental setup as before but centered at 5000\,\AA, and the other at a higher resolution with a $0.46''$ slit width and the 1200 l/mm grating centered at 6010\AA\ with a 10 minute exposure time for an effective resolution of $R=7200$ and binned to a dispersion of $0.6''/$pixel. 

\subsubsection{NIR Spectroscopy}

A near-infrared JHK spectrum was also taken with the FLAMINGOS-2 instrument on Gemini-S in Poor Weather Mode on March 14, 2014. It is shown in Figure \ref{fig:irspecs}. The observations were made with 4 observations of 120 seconds, a 6-pixel slit width, and the HK grism under the HK filter.  The data reduction was performed using the Ureka package provided through Gemini Observatory which contains routines specific to the instrument. Telluric corrections and flux calibrations were performed using the A0V star HD 169257.

A second near-infrared spectrum was obtained with the Magellan FIRE instrument on April 28th, 2015 covering the range from $0.8-2.5\,\mu$m with a spectral resolution of $R=6000$. Data was reduced via the FIRE data pipeline hosted by MIT\footnote{http://www.mit.edu/people/rsimcoe/FIRE/ob\_data.htm}.

\subsection{X-rays}

X-ray observations were made with the {\it Chandra X-ray Observatory} as a part of the Chandra Galactic Bulge Survey \citep[GBS][]{Jonker11,Jonker14}. The GBS consists of many 2\,ks observations covering the Galactic Bulge the twelve square degree region $-3^{\circ}\le l \le 3^{\circ}$, $1^{\circ}\le |b| \le 2^{\circ}$, which purposefully avoids the central Galactic Plane and the accompanying high extinction while preserving the high density of X-ray binaries that are the primary science target of the survey. Observations were made with the I0-I3 CCDs of the {\it Chandra} ACIS-I instrument \citep{Garmire97}. CX330 was observed in Observation ID \#10015, at an off-axis angle of $6.24'$ and a 95\% confidence positional uncertainty $2.5''$ in radius following the methods of \citet{Hong05}. There is some overlap in GBS pointings, but CX330 does not appear on chip in any other observations. The X-ray data is discussed in further detail in \citet{Jonker11}.  Matching X-ray sources to variable optical counterparts is discussed in detail in \citet{Britt14}. 

\section{Multiwavelength Results}
\label{sec:data}

\begin{figure*}
\centering
\includegraphics[width=0.7\textwidth]{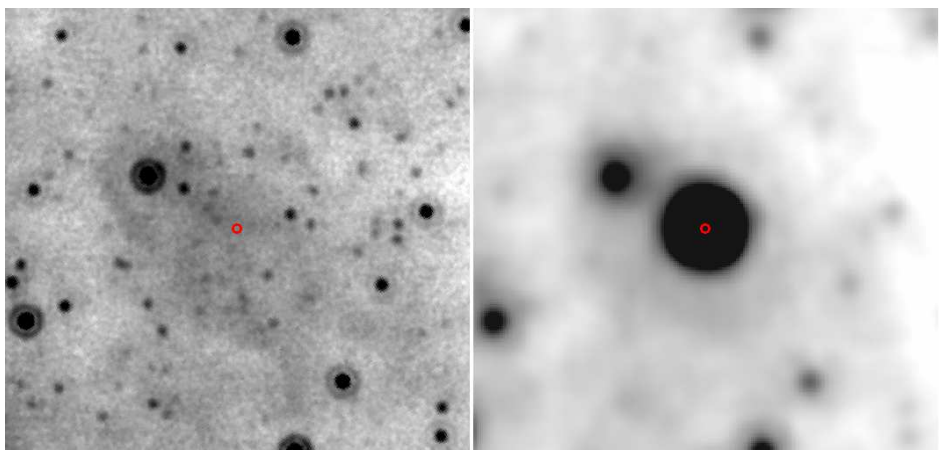}
\caption{A comparison of CX330 brightness before and after outburst. {\bf a:} CX330 before outburst in archival MIPSGAL $24\,\mu$m images, taken on October 1, 2006. 
  The X-ray position for CX330 from Chandra is overlaid in red. North is up and East is left. Each panel is $5'$ across. 
{\bf b:} CX330 in outburst in archival WISE W4. It is visible in all WISE bands,
  $m_{W1}=7.72$, $m_{W2}=6.01$, $m_{W3}=3.32$, and $m_{W4}=2.43$. While not visible in MIPSGAL, CX330 dominates the field in
  W4, which is especially significant considering that the MIPSGAL survey is deeper than WISE. As in the left panel, the X-ray position from Chandra is overlaid in red. 
}
\label{fig:wise}
\end{figure*}

CX330 began an optical and NIR outburst at some time in between 2007 and 2010. The outburst is $>6.2$ magnitudes (i.e. a factor of at least 300) in the mid-IR, as the star is undetected in Spitzer's MIPSGAL $24\,\mu$m band in 2006 but dominates the field in shallower WISE $22\,\mu$m images taken 4 years later (Figure \ref{fig:wise}). A timeline of observations in different wavelengths is shown in Table \ref{table:timeline}. Because this object is variable in both optical and NIR wavelengths (Figure \ref{fig:sed}), observations taken at different epochs should be compared with extreme caution. Even with estimated errors on the order of a factor of several, however, this object exhibits an enormous infrared excess, indicating surrounding dust and gas at a cooler temperature than the central object, $T_{dust}\approx510$\,K (Figure \ref{fig:sed}). 

\subsection{Optical and Near Infrared Photometry Before Outburst}

Prior to X-ray detection, GBS initial imaging showed that this object is not visible down to a limiting magnitude of $r', i'=23$.  There is a star visible inside the Chandra X-ray error circle in the optical and NIR images from before the outburst, but this is at an RA=17:36:43.99 DEC=-28:21:22.45, which is a different position than the transient in optical images located at RA=17:36:43.88 DEC=-28:21:21.30. The true optical/IR counterpart to CX330 is undetected prior to WISE observations in 2010, including in the deeper GLIMPSE ($\lambda_{eff}=3.6\,\mu$m, $4.5\,\mu$m, $5.8\,\mu$m, $8.0\,\mu$m) and MIPSGAL ($\lambda_{eff}=24\,\mu$m, $70\,\mu$m) surveys taken with the {\it Spitzer Space Telescope} \citep{Churchwell09,Carey09}. Limiting magnitudes from pre-outburst photometry are presented in Table \ref{table:timeline}.

\begin{table*}
\begin{center}
\noindent\resizebox{\linewidth}{!}{%
\begin{tabular}{l c c r}
Instrument & Filter & Date & Brightness \\
\hline
USNO-B1.0 & $BRI$ & September 9, 1976 & $>19.6$, $>19.1$, $>17.4$ mag \\
2MASS & $JHK$ & July 2, 1998 & $>14.8$, $>13.8$, $>13.0$ mag \\
Mosaic-II & Sloan $r'$, Sloan $i'$, H$\alpha$ & June 10, 2006 & $>23$ mag \\
MIPSGAL & $24\mu$m & October 1, 2006 & $>8.6$ mag \\ 
GLIMPSE & $3.6\mu$m, $4.5\mu$m  & May 8, 2007 & $>13.5$, $>13.5$ mag \\ 
  & $5.8\mu$m, $8.0\mu$m &  &$>12.5$, $>12$ mag\\
\hline
Chandra ACIS-I & $0.3-8$\,keV & February 4, 2009 & $10^{-13}$\,ergs\,cm$^{-2}$\,s$^{-1}$ \\
WISE & $3.4\mu$m, $4.6\mu$m, $12\mu$m, $22\mu$m & March 15, 2010 & $7.72$, $6.01$, $3.32$, $2.43$ mag \\
OGLE IV & $I$ & April 3, 2010 - July 10, 2012 & $14.6-16.3$ mag \\
Mosaic-II & Sloan $r'$ & July 9, 2010 & 17.2 mag \\
VVV & $ZY$  & Sept 9, 2010 - June 22, 2012 & $14.07$, $13.18$ mag\\ 
    & $JHK$ &  &   $<11.5$, $<11.5$, $10.44$ mag \\
Flamingos-2 & $K$ & March 14, 2014 & - \\ 
Goodman & $BVR$ & April 7, 2015 & Spectrum \\
FIRE & $JHK$ & May 4, 2015 & Spectrum \\
\hline
\end{tabular}
}
\caption{A timeline of all observations of CX330 in chronological order. A black line divides pre- and post-outburst, though it is unclear whether the X-ray observations were made before or after the outburst began. Magnitudes are in the Vega system.}
\label{table:timeline}
\end{center}
\end{table*}

\subsection{Optical and NIR Photometry after Outburst}

In July 2010, a multi-epoch variability survey conducted with the Mosaic-II instrument on the Blanco 4m telescope \citep{Britt14} revealed a variable counterpart with $r'=17.2$. The GBS X-ray observations were taken in between these optical observations on February 4th, 2009. In 2010 observations, CX330 shows aperiodic flickering over a range of 0.3 magnitudes, with the largest intra-night variations of 0.15 magnitudes. The X-ray to optical flux ratio, before correcting for extinction, was estimated to be $F_{X}/F_{r'}=10^{-0.9}$, which drops to $F_{X}/F_{r'}=10^{-2.8}$ with extinction for the line of sight to the Bulge as measured by \citet{Gonzalez12}, and drops further with any local extinction. In DECam observations in 2013, taken over 2 nights with an average sampling of 15 minutes, CX330 also appears at $i'\approx17.45$ as a variable with smooth, aperiodic variability in on the timescale of hours, with intra-night variations of 0.15 magnitudes. 

\begin{figure*}
\begin{center}
\includegraphics[width=\textwidth,angle=0]{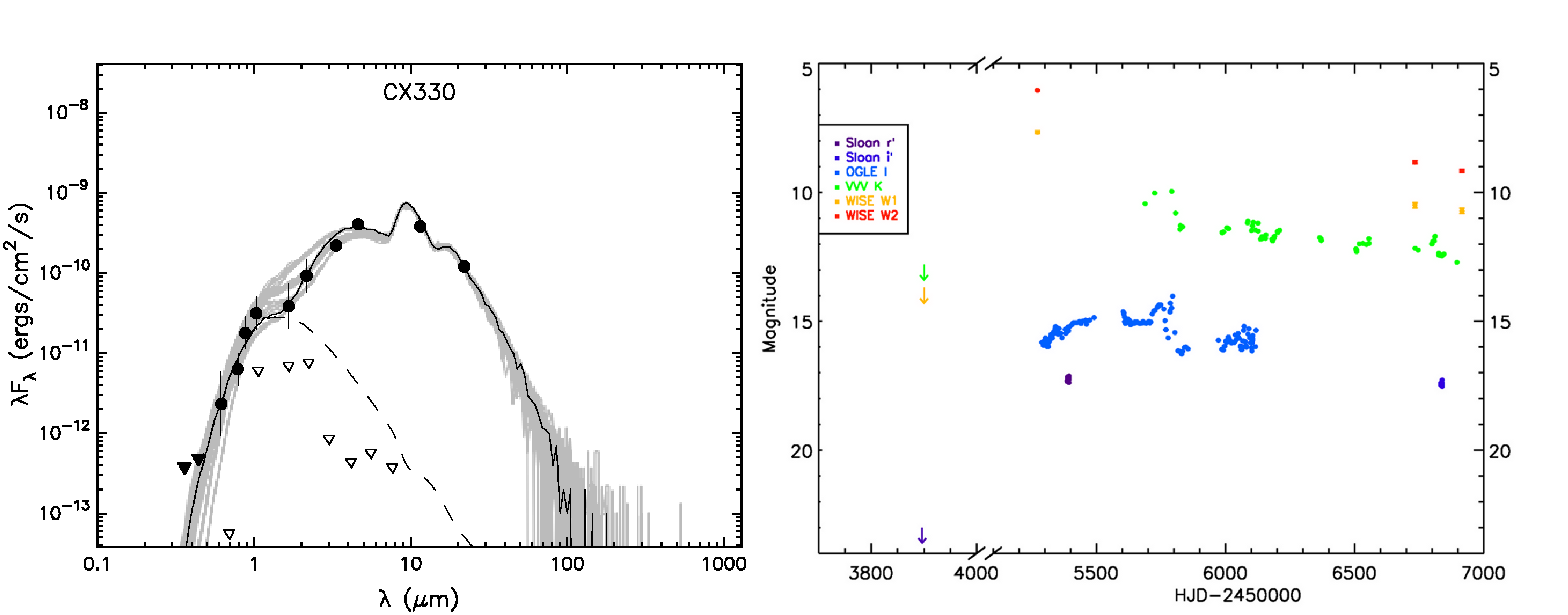} 
\caption{Spectral Energy Distribution and Light curve of CX330. {\bf a:} An SED of CX330 fit with a variety of young stellar object (YSO) models\protect \citep{Robitaille12}. The dotted line represents a reddened single temperature blackbody. Empty triangles are upper limits from photometry before the outburst began. Because CX330 is variable while in outburst, the comparison of data at different epochs is dubious, and we therefore have included error bars that reflect the amplitude of variability observed in each bandpass for observations taken after the WISE data. The modeling clearly demonstrates that a large, cooler dusty envelope around the central hot source is necessary to fit the IR excess.
{\bf b:} Available lightcurves of CX330 since 2010. Upper limits are provided where available.}
\label{fig:sed}
\end{center}
\end{figure*}

OGLE-IV shows CX330 as an irregular variable \citep{Udalski12} with flickering and a very slow decay time over the observational period. The lightcurve, along with Mosaic-II, DECam, VVV, and WISE and NeoWISE observations, is shown in the right-hand panel of Figure \ref{fig:sed}.

CX330 is at least 7 magnitudes brighter in W1 ($\lambda_{eff}=3.6\,\mu$m) than in the GLIMPSE IRAC 3.4\,$\mu$m band, and at least 6 magnitudes brighter in W4 ($\lambda_{eff}=22\,\mu$m) than in MIPSGAL $24\,\mu$m images.

\subsection{Astrometry}

The proper motion and parallax of CX330 were fit independently in the $\alpha \cos{\delta}$ and $\delta$ dimensions using a robust technique involving an iterative reweighting of data points as a function of their residuals. We used the 108 Ks band epochs with seeing $<1.2''$ of VVV tile b361 available to us as of April 30th 2014, a total epoch baseline of 2.9 years. We selected 97 references sources from within $62''$ of the target through an iterative rejection of sources with significant proper motion. We measure a total proper motion relative to the reference sources of $1.6\pm3.3$\,mas\,yr$^{-1}$ and a parallax of $2.8\pm2.5$\,mas.
The proper motion components are $1.2\pm2.6$\,mas\,yr$^{-1}$ and $1.0\pm2.0$\,mas\,yr$^{-1}$ in $\alpha \cos{\delta}$ and $\delta$ respectively, and parallax components are $2.5\pm2.5$\,mas and $44.6\pm28.4$\,mas in $\alpha \cos{\delta}$ and $\delta$. The final parallax is the average of the two measurements weighted by the reciprocal of their uncertainty squared, resulting in a value of $2.8\pm2.5$\,mas. 

\subsection{Optical and Near Infrared Spectroscopy}
\label{specdisc}

The optical spectrum taken in 2014 shows narrow emission lines which we used to measure the radial velocity of CX330. The heliocentric radial velocity we measure to be $60\pm15$\,km\,s$^{-1}$, where the uncertainty is dominated by the wavelength calibration of the spectrum.

The lower resolution spectrum of 2015 covers H$\beta$ but the extinction is high enough to quench it completely, despite the large H$\alpha$ line, while the nearby broad [OIII] doublet at 4959\,\AA\ and 5007\,\AA\ is still present with $F($[O III 5007]$)/ F($H$\beta )>20$, suggesting either very high intrinsic $I($[O III 5007]$)/ I($H$\alpha )$ values or that the [O III] lines are produced in a region separate from the Balmer emission. 

We can use the Balmer decrement to measure the reddening to the source \citep{Osterbrock}: 

\[E(B-V)=\frac{2.5}{k(H\beta)-k(H\alpha)}\log\frac{H\alpha/H\beta_{obs}}{H\alpha/H\beta_{int}}
\]

\noindent where $k(\lambda)$ is the extinction coefficient at that wavelength for a given extinction law. We place a lower limit on the observed ratio $H\alpha/H\beta>280$, which for an intrinsic ratio of $H\alpha/H\beta\approx3$ and taking $k(H\beta)-k(H\alpha)=1.25$ for the Milky Way \citep{Seaton79} gives $E(B-V)>4.1$. Using other extinction laws with lower values of $k(H\beta)-k(H\alpha)$ found in the literature drives the reddening up higher. We use the highest value found in the literature as a conservative lower limit on $E(B-V)$. Even with this conservative assumption, however, this is much higher than the line of sight extinction measured in the VVV survey \citep{Gonzalez12}, $E(B-V)=2.06\pm0.23$ following the extinction law of \citet{Cardelli89}. This leaves $E(B-V)>2$ remaining to local extinction. If this object is in front of the Galactic Bulge, this lower limit moves up, while moving beyond the Bulge in this line of sight adds no substantial foreground reddening as the radius from and height above the center of the Galaxy continue to increase well above a scale height of the disk. We can therefore firmly conclude that the extinction responsible for quenching H$\beta$ is local. 

The [NII] lines in 2015, $>5$ years after the outburst begins, are substantially weakened compared to 2014, with [N II 5755] $\sim 4\times$ fainter than [C IV 5805], when they were of roughly equivalent strength.

\begin{figure*}
\centering
\includegraphics[width=0.7\textwidth,angle=90]{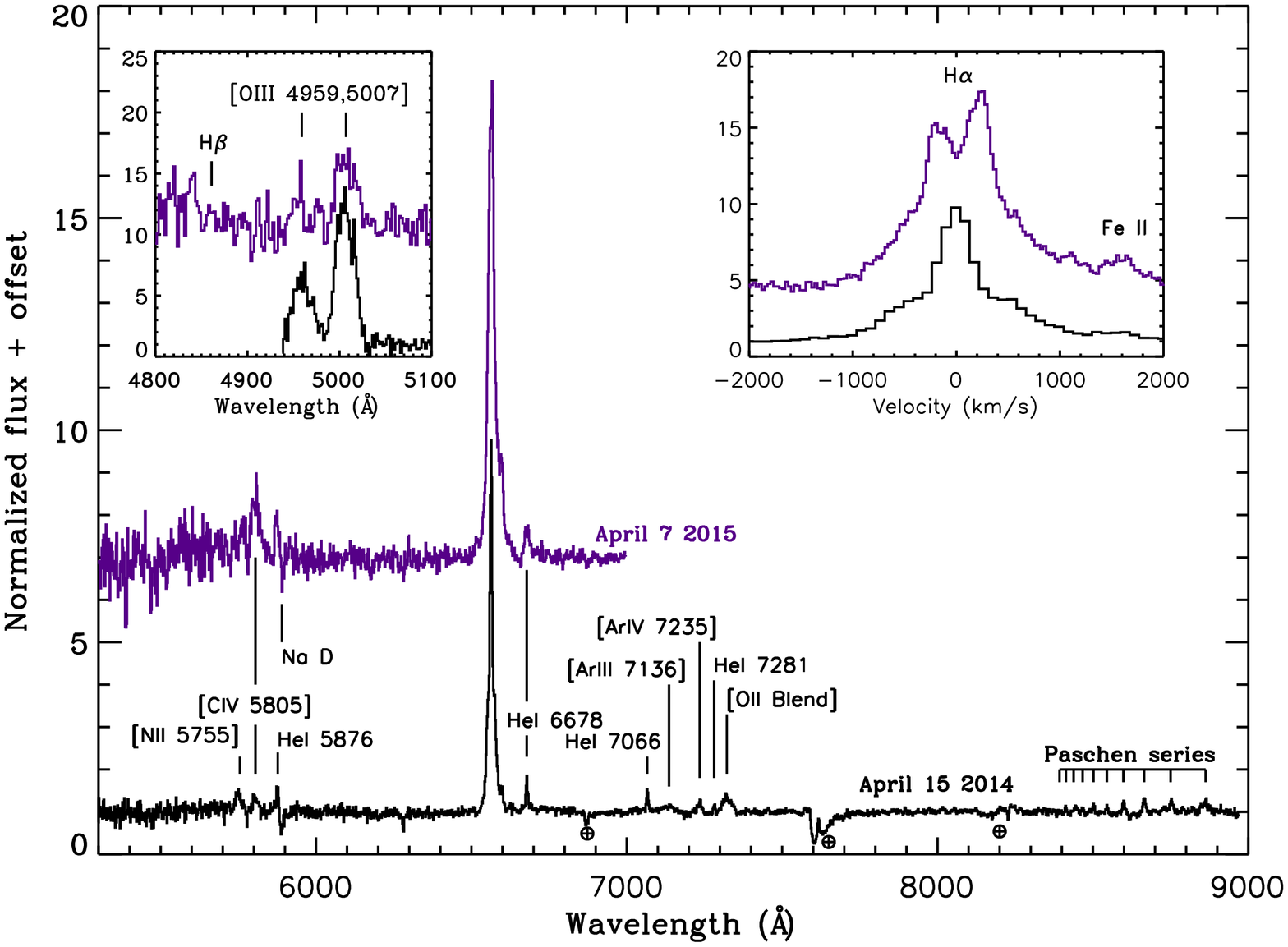} 
\caption{Optical spectroscopy of CX330 in 2014 (black) and 2015 (purple), normalized to continuum with an arbitrary offset added. Telluric lines are marked with $\oplus$. The presence of [O III] is notable because it requires a strong UV ionizing flux above 35.1\,eV to ionize $O^{+}$ and produce $O^{++}$ or comparably high shock temperatures especially since the neutral oxygen feature [O I] at 6300\,\AA\ is absent.  In 2014, the forbidden emission lines are also much broader than the core hydrogen and helium features, though there is a broad pedestal at the base of H and He lines that shares the same velocity profile as the forbidden lines. 
}
\label{fig:optspec}
\end{figure*}

\begin{figure*}
\includegraphics[width=0.7\textwidth,angle=90]{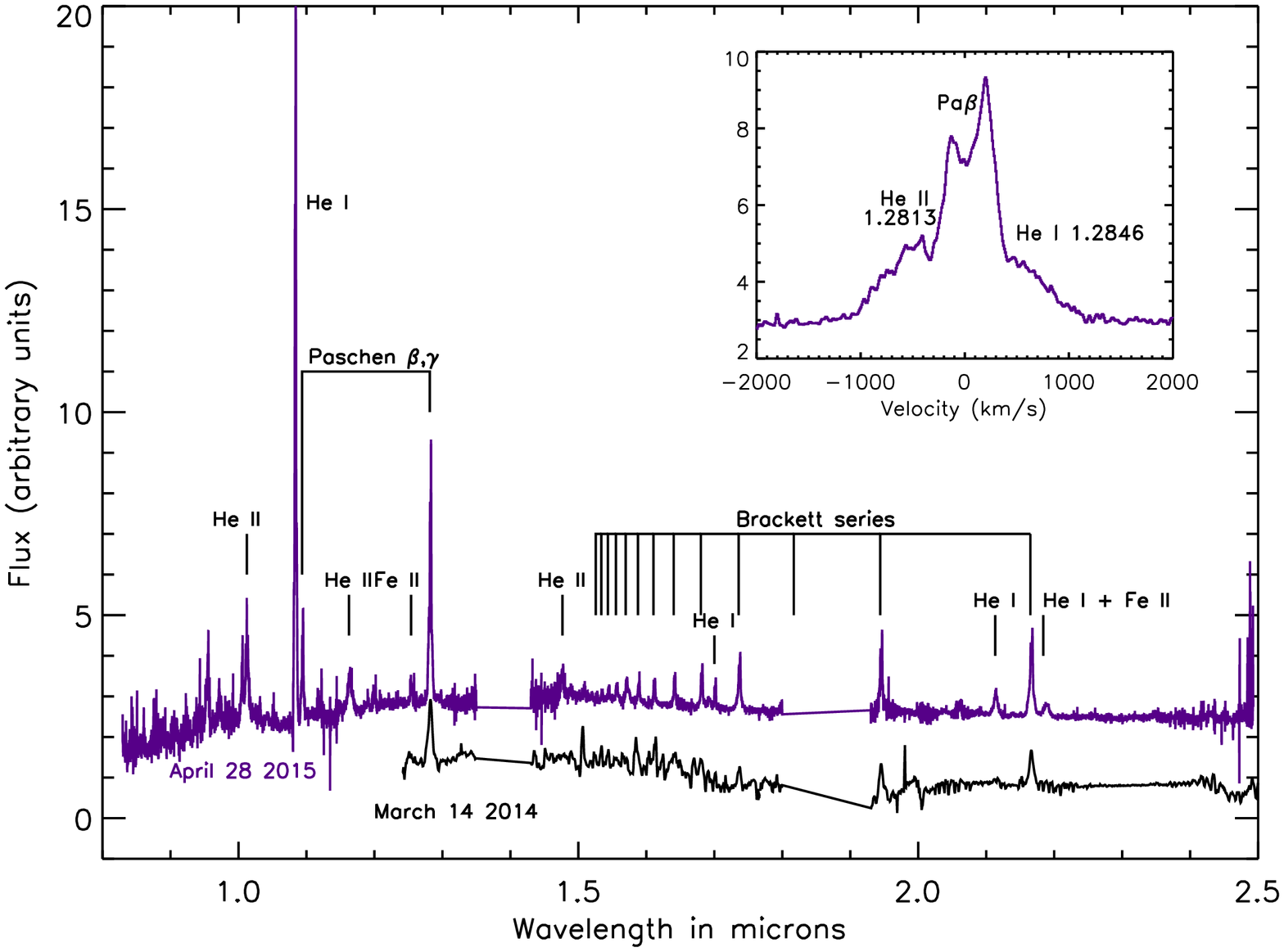} 
\caption{NIR spectra of CX330. The first IR spectrum obtained with FLAMINGOS-2 in March, 2014 is plotted in black. This observation was taken in poor weather, leaving a noisy spectrum. The widths of the spectral lines in this observation are dominated by the low instrumental resolution. A second spectrum obtained with the FIRE instrument on the Magellan telescope in May, 2015 at much higher resolution, plotted in purple, shows many lines of hydrogen and helium in emission, including He II. Many of the stronger hydrogen lines are not isolated, but are blended with lines of He I and He II. 
}
\label{fig:irspecs}
\end{figure*}

Each epoch of optical spectroscopy of CX330 (Figure \ref{fig:optspec}) shows strong hydrogen emission, as well as He I emission. The lines are not redshifted beyond the typical dispersion of stellar velocities in this line of sight, which rules out an extragalactic origin. Also present are strong, high ionization state forbidden lines, most notably [O III] 4959+5007\,\AA\ and [N II] 5755\,\AA. OI 6300\,\AA\ and other OI lines are absent. The forbidden emission lines are very broad, with a full-width-at-half-maximum (FWHM) of $1400\,$km\,s$^{-1}$, which for a spherical outflow would imply an expansion velocity of $700\,$km\,s$^{-1}$. The allowed transitions of hydrogen and helium in 2014 have cores that are much narrower, with a FWHM of $<140$\,km\,s$^{-1}$, with a broad pedestal as wide as the forbidden lines. In 2015, this central emission core fades and leaves a broad base with a double peaked profile with peaks at $v=-170$\,km\,s$^{-1}$ and $v=+260$\,km\,s$^{-1}$ with a core blue-shifted by $25$\,km\,s$^{-1}$ relative to the line center, with the blue peak suppressed compared to the red, a common line profile of bipolar outflows.

The near-infrared spectrum shows strong Brackett and Paschen emission in addition to HeI emission lines. The flux calibrated continuum is poorly fit by a single blackbody with extinction, and a second large cool component is necessary to fit the K-band flux which increases moving to longer wavelengths.

 The FIRE spectrum taken in 2015 shows strong Brackett and Paschen series emission as well as both He I and He II, Fe II, and C IV and C III. The velocity profiles of the helium and hydrogen lines is the same as that of H$\alpha$ in 2015 shown in Figure 3, while the Fe II lines are decidedly more flat topped, with a Full Width at Zero Intensity (FWZI) of $990\,$km\,s$^{-1}$. Some of the hydrogen lines such as Pa$\beta$ and Br$\gamma$ have FWZI measures $\sim2000\,$km\,s$^{-1}$ but are also contaminated with He I and He II lines which are present and are broadening the observed base. The amalgamation of light from distinct shock regions in different radial directions from the source and as [OIII] regions cool and begin producing hydrogen recombination could also artificially broaden the lines. Similar to the 2014 FLAMINGOS-2 NIR spectrum, CO and water lines are also absent in this spectrum.

\subsection{X-ray observations}

We first identified CXOGBS J173643.8-282122 (CX330) as an X-ray source with and X-ray flux $F_{X}\approx 6\times10^{-14}$\,ergs\,cm$^{-2}$\,s$^{-1}$ at a position $2^{\circ}$ above the Galactic Mid-Plane. The initial 2\,ks X-ray observations are described in detail in \citet{Jonker11}. In this observation, CX330 has 8 counts above background, a highly significant point source for Chandra's background in the short exposure time. At an off-axis angle of $6.24'$ the pileup for 8 photons in 2\,ks is negligible. The X-ray luminosity in 2009 is $\approx d_{kpc}^{2}\times2\times10^{30}$\,ergs\,s$^{-1}$.  The optical/infrared counterpart for CX330 is $0.8''$ from the Chandra X-ray position, well within the $95\%$ confidence region for  the X-ray observation \citep{Jonker11,Britt14}. While there are so few counts that a wide variety of models can adequately fit the data, they do offer a few constraints. First, there are no photons below 1 keV. Even with very high extinction, it is difficult to have a very soft X-ray spectrum with kT$<0.5$\,keV such as a Super Soft Source without also having an implausibly high intrinsic X-ray luminosity in order to provide harder photons while killing the softer ones. The lack of photons below 1\,keV gives a range of $N_{H}$ that is consistent with the extreme absorption seen in the optical wavelengths, which is suggestive that the dust around the system was present at time of X-ray observations. We stress, however, that with only 8 photons, many different models can fit the X-ray spectrum and a wide range of values is allowed for photon index and $N_{H}$.

\section{Interpretations of this Transient}
\label{sec:interp}

This object is very unusual in several respects, so we take the opportunity here to rule out some interpretations and reconcile some differences with known classes.

\subsection{AGN} 
A $3\sigma$ upper limit on the redshift of $z<0.0003$ based on the hydrogen emission lines in 2014 means that the source must be quite nearby in cosmological scales, yet no galaxy is resolved in any wavelength. We conclude that CX330 is of Galactic origin. A Tidal Disruption Event around a supermassive black hole (or even around an intermediate mass black hole in a globular cluster) is ruled out for the same reason. 

\subsection{Classical Nova} 
Classical novae exhibit a wide variety of phenomenologies, and require careful consideration. We disfavor this interpretation for the following reasons: 

\subsubsection{Dusty at late times} 

CX330 remains heavily obscured, with a strong IR excess, years after outburst. In a nova scenario where the dust is produced in an expanding shell, as in the nova V1280 Sco \citep{Naito12} or Nova Mon 2012 \citep{Munari13}, the dust expands with the nova shell and results in reddening lessening over time \citep{Hachisu14}. In CX330, the reddening measured from $K-I$ remains constant, or is even growing, over years, which suggests that the local dust is either not associated with the outflow or that dust is somehow being continuously generated from a nova even after the nebular phase has begun. While some evolved stars generate copious amounts of dust, we can firmly rule out a symbiotic nova with a donor from the Asymptotic Giant Branch to most of the Red Giant Branch based on the pre-outburst upper limits even taking into account the large amount of reddening observed. We can hide at a distance of the Galactic Bulge less evolved stars at the start of the RGB only by assuming that all of the copious local reddening above that in the line of sight to the Galactic Bulge ($E(B-V)\ge2$) is present locally before the outburst begins; however, RGB stars typically have mass loss rates from wind from $10^{-9}-10^{-5}\,M_{\odot}\,$yr$^{-1}$, with higher rates only possible for the more evolved, luminous stars which we can rule out firmly, and the others do not produce material in anywhere close to sufficient quantities to provide the observed reddening to this object. 

\subsubsection{The long decay timescale of CX330} 

The outburst of CX330 lasts $>5$ years and is ongoing at the time of writing, decaying only $\sim3$ magnitudes in that time. It is unclear whether or not OGLE-IV observations catch the peak of the outburst. This area is covered by ASAS down to $I=14$ until Oct. 26, 2009, so an outburst peak could be hidden while CX330 is behind the Sun from November, 2009 - February, 2010. Most classical novae dim much faster from peak than CX330 does (assuming that the peak is observed), decaying several magnitudes on a timescale of weeks to months \citep{Strope10}, originating from much smaller spatial scales than those in YSOs. Even if the peak of the outburst is unobserved, this lightcurve remains unusually slow for a nova. Some rare classical novae, however, can proceed much more slowly and form large amounts of dust which regulates the brightness of the novae through the optical depth \citep{Chesneau08}. Objects like V1280 Sco have a long plateau phase after dust production that lasts years, but there is no high frequency variability in this phase \citep{Naito12} while observations of CX330 with DECam in 2014 show rapid variability on the time scale of a day or less even 4 years after the outburst begins and while its SED peaks in the near-infrared. Also, there is a fairly well defined relationship between the outflow speed of a classical nova, $v_{out}$, and the decay speed, $t_3$ \citep{Esenoglu00}. While novae can have a variety of lightcurve morphologies \citep{Strope10}, the scatter this may introduce into the relationship between $v_{out}$ and $t_3$ is much less than the deviation of CX330 in comparison if we assume that the peak of the outburst is observed in OGLE-IV (see Figure 3 of Esenoglu et al. 2000). Not only do no classical novae known have $t_3$ as long as CX330's now is, 2100 days measured in $I$ band, but those few that do have $t_3>1$ yr have outflow speeds of only a few tens of km\,s$^{-1}$ \citep{Warner03}. This argument only applies to the case that we observe the outburst peak, otherwise we have no starting point with which to measure $t_3$.

\subsubsection{Nova returning from dust dip requires fine tuning} 

One possibility considered that avoids the problem of $t_{3}$ versus $v_{out}$ discussed above is that the outburst we observe is not the initial nova, but the recovery from a dust dip as in D class novae \citep{Strope10}. However, in this case the eruption time and duration of a classical novae must be fine tuned to avoid detection in all-sky surveys.

In D class novae, the dust dip lasts for $100-200$ days, with those novae generating more dust having longer dips. If the small rise seen at the start of OGLE observations is a recovery from a dust dip, then the nova eruption must have occurred in the $100-200$ days prior to WISE observations on March 15, 2010, likely closer to 200 days given the large amount of reddening observed. The gap between ASAS and OGLE observations leaves 141 days for CX330 to nova while it is behind the Sun and return from a dust dip. This is not a short enough time to absolutely rule out a D class nova eruption on its own, but it does require that the nova go off within a month of ASAS observations stopping and that the duration is unusually short for the amount of dust generated. 

\subsubsection{Variability at late times} 

The variability at late times seen in CX330 is uncharacteristic of D class nova. The $I$ band lightcurve in OGLE IV is smooth for over a year, with a possible rebrightening while behind the sun between MJD 55500 and 55600. There are observed rebrightenings lasting $1-2$ months in the second year of observations, which then remains flat until halfway through the 3rd year of observation when variability on timescales of days begins.

It is not uncommon for novae to rebrighten after the initial eruption \citep{Strope10}. These ``jitters'', however, follow similar patterns; they start before the nova enters the nebular phase and the time between them grows logarithmically with time since the start of the outburst \citep{Pejcha09}. Because they are likely caused by sudden changes in the hydrogen burning rate in a white dwarf envelope, they occur on massive white dwarves in fast novae \citep{Pejcha09}. Any slow nova, as late dusty novae must be \citep{Williams13}, must originate on a low mass white dwarf. The rebrightenings seen in fast novae are therefore not a viable explanation for the late time variability.

\subsubsection{Estimate of ejecta mass required} 

In the finely timed scenario of a D-class dust-dip nova discussed above \citep{Strope10}, we can use the lower limits on reddening along with the outflow speed and time since eruption to make a crude estimate of the mass of the ejecta. 

For a nova in the Bulge at 10\,kpc experiencing the full amount of interstellar reddening \citep{Gonzalez12}, the remaining local extinction amounts to $E(B-V)>2.0\pm0.6$. Assuming average interstellar gas-to-dust ratios, the column density from local sources is $N_{H}=5.6\times10^{21}$\,atoms\,cm$^{-2}\times {\rm E(B-V)}>1.1(\pm0.3)\times10^{22}$\,atoms\,cm$^{-2}$. In a nova scenario where the absorbing material is generated in an outflow started at the time of the eruption, the material is at a radius of $R\approx1000$\,AU from the nova after 5 years given the observed velocity widths. Assuming that all of the Balmer emission is behind all of the mass as a lower limit to the amount of mass in the ejecta required to achieve the reddening seen, we still require $\approx0.01$\,M$_{\odot}$ to generate the required column density 5 years after an outburst moving at the observed speed. This is orders of magnitude higher than the most massive nova ejecta, which themselves require low accretion rates and therefore occur very infrequently \citep{Yaron05}. We stress again that the limit on local $E(B-V)$ from spectra is conservative in two respects: H$\beta$ is assumed to be just below the detection threshhold when it may be fainter and we use the most conservative extinction law we could find \citep{Seaton79}. It seems certain, therefore, that a single nova event is incapable of producing the observed amount of dust, while dust from past novae should be ejected by the binary before recurrence.

\subsubsection{Spectral evolution} 

At first glance, the emission line profile in 2015 is a much narrower version of the P Cygni profiles seen in the nova V1280 Sco only 2 weeks after the eruption \citep{Das08} or the spectrum of the nova KT Eri taken only 17 days after eruption \citep{Munari14}. The high order Paschen and Brackett lines are very strong, meaning the density is high in the region where these lines are being produced. 

However, this spectrum was taken $>4$ years after the outburst began. At the expansion velocity of the forbidden lines, a nova shell should have expanded $\gae600$\,AU, which makes it difficult to envisage any nova scenario maintaining a sufficient density to explain the strong Paschen lines \citep{Raj15}.
The late epoch spectra resemble some novae at early times, yet the color of CX330 remains roughly constant, or even reddens, with time (Figure \ref{fig:sed}). In dust producing novae, the broadband colors generally become bluer with time as the optical depth of dust drops over time \citep{Hachisu14}. We conclude from the constant (or slight reddening) color that the local dust cloud around CX330 is not expanding, but may be cooling. 

\subsubsection{Helium Nova}

Helium novae (fusion of helium on the surface of a white dwarf primary) can be very red and long lasting, but are hydrogen deficient by definition \citep{Woudt05}. CX330 shows very strong hydrogen lines, ruling out this interpretation.

\subsubsection{Symbiotic Recurrent Nova}

Symbiotic novae can last years or decades, with dust that is not as quickly ejected from the wider system. While this could explain the constant reddening seen in CX330, dust still needs to be local to the system before outburst to hide the donor star, and the donor star must still be low on the RGB. Since winds from such stars are inadequate to produce this dust, we here consider a recurrent nova (RN) in which the ejecta do not have time to fully dissipate before another eruption renews the dust. While there is no evidence of a giant companion in the spectra, it is clear from the magnitude of the outburst that any companion present cannot contribute a significant portion of the light at any wavelength. However, RNe require high mass white dwarfs in order to trigger a nova with less material on the surface. To produce $\gae10^{-2}\,M_{\odot}$ of dust around the binary when RNe eject on the order of $10^{-5}-10^{-6}\,M_{\odot}$ in an eruption requires the material of $\gae10^{3-4}$ eruptions over $\gae10^{4-6}$ years.  We consider that the dust from novae many thousands of years ago cannot remain local enough either to be heated to $>500K$ as suggested by WISE colors or to provide the optical depth observed.

\subsection{Luminous Red Nova} 

This class has become much better understood in the last 10 years and is likely a misnomer unrelated to classical novae. The favored interpretation is that they are the result of stellar mergers \citep{Tylenda11}. Though the infrared emission can last longer, the optical light, with some exceptions, fades in a matter of weeks to months, much faster than is observed in CX330. More importantly, the absolute magnitudes of these events are quite bright, and would place CX330 well beyond the Bulge either in the Galactic halo or even outside of the Galaxy. Several progenitor scenarios suggest that they have giants as precursers, whereas CX330 is undetected in deep observations in optical and IR wavelengths before the outburst began. Also, a giant precursor makes the large outflow velocity much harder to achieve since they have much lower escape velocities than main sequence stars. For mergers of main sequence stars, the timescales are much shorter than those for giants, e.g. the merging contact binary V1309 Sco \citep{Tylenda11}, and makes the length of the outburst difficult to achieve. 

Perhaps a region of parameter space exists where a merger between a subgiant and brown dwarf companion could be long lived and faint compared to the usual transient types. This scenario requires careful simulation to totally eliminate, but seems fine tuned.

\subsection{Accretion Driven Outburst in a YSO}

\subsubsection{Spectroscopy}

The optical and infrared spectra are unusual for YSOs in outburst, consisting of emission lines on a hot, featureless continuum. As discussed in Section \ref{specdisc}, the nebular emission lines are broad in each observation, while the allowed hydrogen lines transition from being narrow and single-peaked with a broad base to broad and double-peaked, with the blue peak suppressed relative to the red. This is a profile commonly seen in bipolar outflows \citep{Steele11}. The separation between peaks observed in the H$\alpha$ profile in 2015 is an order of magnitude higher than one would expect from a Keplerian disk at AU scales. In the YSO interpretation, then, we will more closely examine the case that a bipolar outflow is responsible for the double-peaked feature as a Keplerian disk does not fit.
Bipolar outflows are a common feature of accreting systems, and are limited in velocity to roughly the escape velocity of the region from which they are ejected \citep{Livio99}. If we assume an accretion rate comparable to the most extreme YSO transients, FU Orionis objects (FUors), and a compactness corresponding an escape velocity of the outflow speed observed, then it is clear that the accretion temperature is not hot enough to photoionize [OIII]. 

Another way to produce high ionizations is through shocks, which occur when fast moving material encounters slower material and collides \citep{Osterbrock}. The shock temperatures corresponding to the observed outflow velocities are high enough to produce these lines \citep{Osterbrock}. We therefore find that [OIII] and other high ionization lines in this scenario are produced in a shock from an outflow, from a wind or jet plowing into what is left of the cooler envelope of gas feeding the star formation. A central object massive enough to generate [OIII] or He II through photoionization should be visible in pre-outburst photometry even at Bulge distances, as discussed in further detail below.  We envisage that the high ionization states are a result of this shock rather than from photoionization from the central source. The velocities observed are higher than in most other YSOs and can be explained by having a more compact or more massive central object.  The higher velocity outflow then gives a hotter shock region. As the shock propagates and cools, hydrogen can recombine which would explain the broader line profiles in recombination lines in 2015 compared to 2014 as the shock regions begin to dominate the emission. Since shocks can be produced in many locations which we do not resolve, the velocity distribution may be artificially broadened by thermalized gas shocking in different radial directions from the central source.

Lithium absorbtion at $\lambda\lambda=6707$\AA\ is a common feature of YSOs, though our observations are of too low a signal to noise ratio to expect to detect this line. Our $1\sigma$ upper limit for the equivalent width of Li 6707 is EW$<0.2$, which is not discriminating for YSOs, for which EW$=0.1-0.2$\,\AA\ is not unusual. This also does not account for shielding of the line by an elevated continuum, which must be present in CX330 since the continuum at time of spectroscopy is at least 300 times brighter than the base level of the photosphere. 

The X-ray spectrum, such as it is, is consistent with that of eruptive YSOs such as FU Orionis \citep{Skinner10}, which consist of two components, one hard ($kT\approx 3-4$\,keV) and one soft ($kT<1$\,keV) thermal component with high absorption ($N_{H}\approx 10^{23}$), though with only 8 photons we stress that many other models are also consistent with the available X-ray data. 

The absence of absorption lines in any part of the spectrum, particularly the CO lines at $2.3\,\mu$m, is inconsistent with the previously largest class of YSO outbursts, the FU Orionis (FUor) class, but could be explained with a hotter accretion disk than is seen in those objects. The FUor class has been described using a small number of objects, and are only one manifestation of the broader class of ``eruptive YSOs'', which includes objects that sometimes do not show CO absorption. Indeed, as the sample of eruptive YSOs grows, it is becoming clearer that many objects do not fit neatly into the FU Orionis or smaller, shorter EX Orionis class, but contain some characteristics of both \citep{Contreras16b}. It is also important to note that we do not detect the stellar photosphere in any observations; the outburst is so large that the light is completely dominated by some combination of the accretion disk, reprocessing from the dusty envelope around the system, and shocks associated with the outflow. This makes it extremely difficult to precisely constrain the stellar properties. 

\subsubsection{Isolation and Presence of Interstellar Dust}

In the YSO interpretation, the most interesting characteristic of CX330 is its position on the sky. It is $2^{\circ}$ above the Galactic Plane, and well outside of any known star forming region. All known large magnitude outbursting YSOs such as FUors are located in star forming regions, likely in part because of selection biases but the actual distribution is also concentrated in these regions \citep{Contreras14,Contreras16a}.
Additionally, CX330 must have formed in situ, since its measured proper motion with the VVV survey has an upper limit of $3.3$\,mas\,yr$^{-1}$. The phase of stellar evolution in which disk instability events occur is generally thought to be within $10^{6}$ years after the formation of the disk \citep{Hartmann98} so that it can have drifted less than a degree since the time of formation.
Upper limits on the parallax angle from VVV astrometry place a lower limit on the distance of 350 pc.
Distances near the Galactic 
Bulge, on the other hand, are also unlikely because then that would imply an 
X-ray luminosity of $L_{X}\approx 5\times10^{32}\,{\rm ergs \, s^{-1}}$,
two orders of magnitude more luminous than FU Orionis itself, which has $L_{X}\approx 6\times10^{30}\,{\rm ergs \, s^{-1}}$  \citep{Skinner10}.  A distance
of $\sim 1-3$ kpc is consistent with the X-ray brightness observed, the lack of observed proper motion, and the
lack of optical or IR detection prior to outburst. A height of $2^{\circ}$ above the disk at 2\,kpc translates to a height above the disk of 36\,pc, which is within the scale height of gas for the Milky Way which is $\sim 100$\,pc.

\begin{figure}
\includegraphics[width=0.5\textwidth,angle=0]{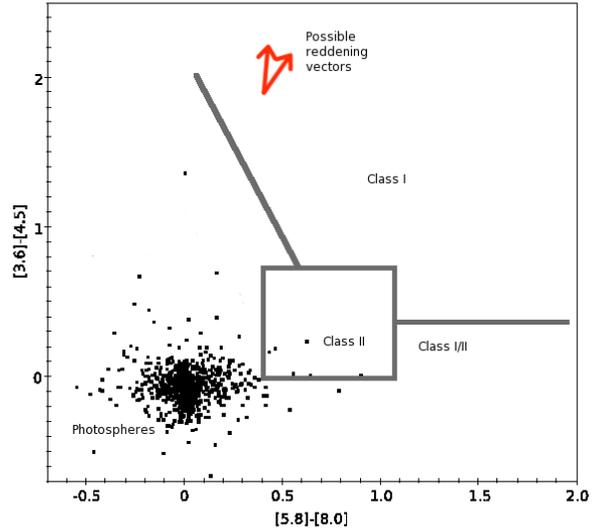} 
\caption{The {\it Spitzer} GLIMPSE point source catalog contains no population of protostars within $2'$ of CX330. Using criterion to select YSOs in the IRAC color-color map \citep[e.g.][]{Robitaille08,Qiu08}, we find no evidence of a cluster of star formation in this region. Before outburst, CX330 was not detectable in GLIMPSE limits, which could be true of most stars later than A0 spectral type at most likely distances and all O stars at any distance.}
\label{fig:ccplot}
\end{figure}

Because the progenitor of CX330 is not seen in observations prior to its outburst, it is entirely possible that other intermediate to low mass stars exist in its immediate vicinity. We can place upper limits on the stars around it from existing catalogs and our own data from before the outburst began. All OV and B0V should be detected all the way to the Galactic Bulge with $E(B-V)=4$ (the line of sight reddening is only $E(B-V)=2$ at the Bulge, but we include a substantial fudge factor for local dust that may shroud other systems). At a distance of 3 kpc, which includes 2 spiral arms, we would expect to see all OV and BV stars associated with CX330. For a distance of 1 kpc, we would also expect to see early A stars. If reddening is less than $E(B-V)=4$ for associated stars then cooler objects would also be visible.  Using archival Spitzer point source catalog within $2'$ of CX330, no obvious young stars are visible, as shown in Figure \ref{fig:ccplot}. It seems likely, therefore, that no high mass star formation is taking place associated with CX330. While CX330 cannot be placed on Figure \ref{fig:ccplot} directly since it does not appear in the GLIMPSE catalog, using the WISE colors ($W1-W2=1.7$, $W2-W3=2.7$) and the classification scheme of \citet{Koenig14} places CX330 on the border between Class I and Class II YSOs.
It is unsurprising that intermediate and low mass YSOs in the region are not visible in GLIMPSE catalogs as CX330 itself was below detection limits prior to outburst.

\begin{figure}
\centering
\includegraphics[width=0.45\textwidth]{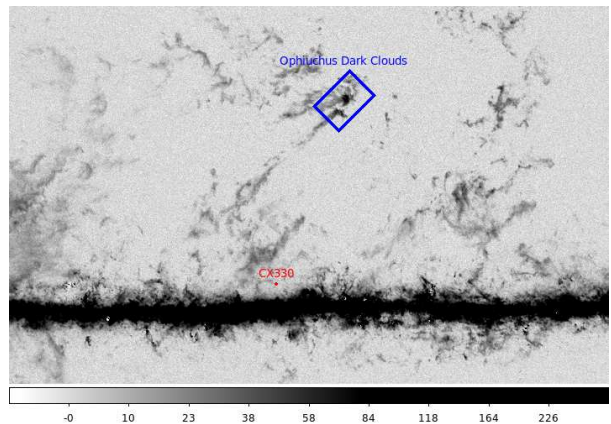}
\caption{The type 3 CO map of the Galaxy produced by the Planck satellite shows that there is no prominent star formation region at this location. A nearby region of star formation, the $\rho$ Ophiuci cloud complex, is highlighted to aid comparison of CO around CX330 with that around star forming regions. For scale, the distance of CX330 above the Galactic Plane is $2.0^{\circ}$.}
\label{fig:planck}
\end{figure}

To estimate the amount of gas and dust in the region we use the dust maps from the {\it Planck} mission \citep[Figure \ref{fig:planck}][]{Planck14}. These yield a column density of molecular hydrogen of $5\times10^{20}$\,cm$^{-2}$, which is $\sim1000$ times lower than the column densities around the locations of known large outbursts of YSOs and $\sim10-20$ times lower than the column density at which a cloud should become unstable to gravitational collapse  \citep{Andre10}. CX330 therefore cannot reside in a cloud as large as a pixel of the Planck maps on the sky ($\sim3'$). The typical angular size of a Bok Globule at 1 kpc would be of order $1'$, while larger, more diffuse complexes would typically occupy a few square degrees. No such globule is seen in optical or NIR images of the area, though there is diffuse dust typical of lines of sight at this latitude. In $24\,\mu$m WISE and MIPSGAL images, this dust is warm and glowing, though this does not mean that the same would have been true $10^{6}$ years ago. CX330 appears within $2''$ of the geometric center of the warm dust which is roughly $2'\times1'$ in area, though CX330 lies at the western edge of the area of higher optical extinction. There may be no real association between CX330 and this dust cloud, but this possibility merits further study. As star formation occurs on the order of dynamical times \citep{Elmegreen00}, faster at small scales, the rapid collapse of a turbulent cloud of gas could proceed within $10^{6}$ years for a small enough cloud.

The ``Handbook of Star Forming Regions'' lists no star forming regions within 3 degrees of CX330. The nearest is NGC 6383, which is at $l=355.7$, $b=0.0$, making it about 4 degrees away. 
The most prominent regions in this direction are the Pipe Nebula and the B59 star-forming core ($\sim8^{\circ}$ away) and the Corona Australis region ($\sim17^{\circ}$ away). Both of these regions are relatively nearby, at distances $\approx130$ pc, making them unlikely to be associated with this object. The latter region is also quite isolated and offset from other molecular material. Beyond those structures, the majority of star forming regions at these Galactic longitudes are located in the Scutum-Centaurus spiral arm of our galaxy, which is about 2-3 kpc from us. There are many prominent HII regions and star forming regions in this arm, including the Eagle Nebula and the Lagoon Nebula (and NGC 6383 above), but there is a notable absence of such regions in the vicinity of CX330.

The CO map presented in Figure \ref{fig:planck} is the Type 3 CO Discovery map produced by the Planck satellite, which combines CO line ratios and models of the CMB, CO emission, synchrotron and free-free emission to identify faint molecular clouds \citep{Planck14} with a resolution of $5'.5$. In order to ascertain how isolated CX330 is compared to YSOs with large outbursts, we compare the velocity-integrated CO distribution of the sky within $30'$ of CX330 to that near 8 known FUors (they being the most extreme class of outbursting YSO yet found), all associated with the Giant Molecular Clouds of Cygnus or Orion. Every FUor, without exception, has an associated region of dense CO gas well above the Planck background reaching anywhere from $35-100$\,$K_{\rm RJ}$\,km\,s$^{-1}$, while the most CO around CX330 is $15$\,$K_{\rm RJ}$\,km\,s$^{-1}$. The CO concentration at the position of CX330 is also much less than at known FUor stars at 7\,$K_{\rm RJ}$\,km\,s$^{-1}$.
The closest any FUor gets is FU Orionis itself at 13\,$K_{\rm RJ}$\,km\,s$^{-1}$, though it lies at the edge of a much denser region while CX330 does not.

 At least one other flaring YSO has been detected outside of a star forming region, GPSV15 \citep{Contreras14}, though it is unclear how old this object is or if it is still accreting material.

The neighborhood around CX330 in the WISE catalogs reveals only one other object with an IR excess, which is not as red as CX330 and is $5'$ away. There are a few objects with red W2-W3 adjacent to CX330, but these are not visible in images and are likely due to the wings of the profile of CX330 in W3, which is quite spatially extended. This may not indicate a lack of nearby low mass YSOs, however, as CX330 was itself undetectable outside of outburst. In WISE images, only warm dust is visible which could argue for formation in a loose association through turbulence rather than a compact cluster feeding competitive accretion. 

Some relatively advanced young stars ($\sim10^{6}$ years of age) have been seen in isolation but it is not known whether they have formed in situ or if they have drifted from their place of birth \citep{Feigelson96,Contreras14}. While ``isolated'' young stars have been reported in the past, they are associated with large molecular clouds \citep{Grinin91,The94}, while the object reported herein is truly isolated; the observational limits on the proper motion and limited lifetime of this phase of formation mean that it must have formed within 1$^{\circ}$\ of its present location on the sky.  

\subsection{Summary of Interpretations}
YSO outbursts are the only class of variable that explains the presence of warm, non-expanding dust local to the system, absence of a progenitor in deep optical and IR imaging prior to outburst, X-ray emission, extremely long duration of the transient event, and outflow velocities on the order of several hundred km\,s$^{-1}$ in a Galactic object. The presence of a non-expanding cloud of warm dust is naturally explained by a YSO, but is difficult to explain with a nova interpretation, as any scenario involving the white dwarf as the source of the dust must have an expanding dust cloud in order to conceal the donor star, while any donor star capable of producing dust is impossible to reconcile with an absence in pre-outburst imaging without placing the system far beyond the Galactic Bulge. 

It is also possible that CX330 is the first in a new class of object unrepresented by the cases outlined above. 

\section{Discussion}
\label{sec:disc}

The most extreme observed YSO outbursts are known as FU Orionis stars (FUors). All known FUors are located in star forming regions \citep{Hartmann87,Bell95}. Only $8-10$ FUors have been observed to go into outburst, and efforts to search for YSO outbursts outside of star-forming regions have so far found that these outbursts are concentrated in star forming regions \citep{Contreras14}. In large YSO outbursts such as these, it is common to not see photospheric lines as the star's photosphere is several magnitudes fainter than the continuum emission during the outburst.

Simulations of Star Formation (SF) processes focus primarily on environments associated with Giant Molecular Clouds (GMC) because most of the gas in the galaxy is in such clouds ($>80\%$ in clouds with $M>10^{5}\,M_{\odot}$) \citep{Stark06}. SF theories can be grouped into 2 camps: hierarchical SF and clustered SF. In clustered SF, high mass stars are formed through competitive accretion only in regions of very high density, with birth masses lower than the typical stellar mass and increased though accretion of unbound gas in the environment \citep{Bonnell01,Lada03}. Hierarchical SF posits that stars form in a fractal distribution on a smoothly varying range of scales dominated by turbulence rather than magnetic support \citep{Larson94,Bastian07}. There is evidence of hierarchical SF as a result of supernovae shocks \citep{Oey05} and in at least one association of massive stars in a larger SF region \citep{Wright14}. If star formation proceeds in a fractal distribution of scales, some few stars should be observed to form even in total isolation from the cloud complexes which have been the focus of large star formation studies.

There are observational characteristics of CX330 that deviate from known YSO outbursts. Two infrared spectra taken in 2014 and 2015 show no sign of CO $\nu'-\nu''$ 2-0 or 3-1 bands in either absorption or emission at $2.3\,\mu$m. CO in absorption is a defining marker of FU Orionis variables, and is common in low mass YSOs in either absorption or emission \citep{Scoville79,Krotkov80}. It is important to note that CO is destroyed at temperatures above $5000\,$K and collisional excitation requires that the temperature and density be above critical values of $3000$\,K and $10^{10}\,$cm$^{-3}$, respectively \citep{Scoville80}. Additionally, the outflow speeds observed are unprecedented among outbursting YSOs, which typically have an outflow speed $\lae 300\,$km\,s$^{-1}$. Both of these differences can be explained if the gravitational potential is deeper than in previously observed outbursting YSOs. A forming YSO is larger than a Main Sequence star of the same mass. The Main Sequence structure of a star with an escape velocity of $700\,$km\,s$^{-1}$ is therefore a rough lower limit on the mass of CX330, which is approximately $1.5\,M_{\odot}$ \citep{Padmanabhan}. 

Typical YSOs with large outbursts have a reflection nebula visible in the optical or NIR
from the light reflecting off the surrounding molecular cloud. CX330 has 
no such reflection nebula, likely because it is not embedded in a molecular 
cloud of sufficient mass or size to have been detected in optical images.
The mid-IR excess
of CX330 that only appears after outburst is consistent with the reprocessing of
optical light by a dusty envelope in YSOs, so it is likely that
there is an extended local structure of dust and gas that is feeding the accretion in CX330. Indeed, the SED (Figure \ref{fig:sed}) is consistent with class I/II YSOs.

YSOs undergoing extreme accretion disk instability episodes such as FU Orionis types are young enough ($<10^{6}$ years) that a large molecular cloud would not have had time to disperse \citep{Blaauw64}. This indicates that the cloud collapse for CX330 was rapid, favoring turbulent formation scenarios hinging upon the Jean's mass \citep{Larson94}.
As stars undergoing such accretion events are expected to be very young and are predicted to be a repeated but rare phase in most YSO development tracks, this discovery may force a revision of our understanding of typical star formation environments.
Using the large IR outbursts to find new outbursting YSOs offers a new window into star formation theories by sampling a dramatically different phase space of formation conditions than is present in large star forming regions \citep{Contreras14}.

\noindent{\large\bf Acknowledgements} 
\noindent T.J.M. thanks Rob Fender, Selma de Mink, and Anna Scaife for discussions. C.T.B., T.J.M., and J.D.G. thank Neal Evans for useful discussions. C.T.B. thanks Paul Sell for discussions on emission from shocks. C.T.B., T.J.M., and L.C. thank Ulisse Munari for insights on the optical spectral properties. R.I.H. acknowledges support from the National Science Foundation under Grant No. AST-0 908789. P.G.J. and M.A.P.T. acknowledge support from the Netherlands Organisation for Scientific Research. This research has made use of NASA's Astrophysics Data System Bibliographic Services and of SAOImage DS9, developed by Smithsonian Astrophysical Observatory.

\bibliographystyle{mn2e}
\bibliography{cx330}

\end{document}